\documentclass[a4paper,doublecol]{epl2} 

\usepackage{graphicx}
\usepackage{amsmath}
\usepackage{amssymb}
\usepackage{rotating}

\newcommand{\ket}[1]{| #1 \rangle}

\newcommand{\degree}{^{\circ}}
\newcommand{\proj}[2][]{| #2 \rangle #1 \langle #2 |}
\newcommand{\bracket}[2]{\langle #1 | #2 \rangle}
\newcommand{\tr}[1]{\mathrm{tr}\bigl\{ #1 \bigr\}}
\newcommand{\expect}[1]{\langle #1\rangle}

\makeatletter
\renewcommand{\epl@stylemark}{\relax}
\renewcommand{\@oddfoot}{\relax}
\renewcommand{\@evenfoot}{\relax}
\renewcommand{\ps@eplfirst}{%
\renewcommand{\@oddfoot}{\hfil\arabic{page}\hfil}
\renewcommand{\@evenfoot}{\relax}
\renewcommand{\@oddhead}{Posted on the arXiv on 28 October 2009\hfill}
\renewcommand{\@evenhead}{\relax}
}
\makeatother

\begin{document}

\title{Efficient Quantum Key Distribution With Trines\newline 
Of Reference-Frame-Free Qubits}
\shorttitle{Efficient quantum key distribution with trines %
of reference-frame-free qubits\hfill\arabic{page}}

\author{Gelo Tabia\inst{1,2} \and Berthold-Georg Englert\inst{3,4}} 
\shortauthor{\arabic{page}\hfill G. Tabia and B.-G. Englert}

\institute{%
\inst{1}Department of Physics and Astronomy, University of Waterloo, Waterloo, %
Ontario, Canada N2L 3G1\\ 
\inst{2}Perimeter Institute, Waterloo, Ontario, Canada N2L 2Y5\\
\inst{3}Centre for Quantum Technologies, National University of Singapore, 
Singapore 117543\\
\inst{4}Department of Physics, National University of Singapore, 
Singapore 117542
}

\abstract{%
We propose a rotationally-invariant quantum key distribution scheme
that uses a pair of orthogonal qubit trines, realized as mixed states of three
physical qubits. 
The measurement outcomes do not depend on how Alice and Bob choose their
individual reference frames. 
The efficient key generation by two-way communication produces two independent 
raw keys, a bit key and a trit key. 
For a noiseless channel, Alice and Bob get a total of $0.573$ key bits per
trine state sent (98\% of the Shannon limit).
This exceeds by a considerable amount the yield of standard trine schemes, 
which ideally attain half a key bit per trine state.
Eavesdropping introduces an $\epsilon$-fraction of unbiased noise, ensured by
twirling if necessary. 
The security analysis reveals an asymmetry in Eve's conditioned ancillas for
Alice and Bob resulting from their inequivalent roles in the key generation. 
Upon simplifying the analysis by a plausible symmetry assumption, 
we find that a secret key can be generated if the noise is below the 
threshold set by $\epsilon = 0.197$.
}

\maketitle

\section{Introduction}
Mutually non-orthogonal quantum states are important in quantum key
distribution (QKD) because such states cannot be completely distinguished 
from each other and hence they are intentionally used to transmit classical
information while preventing eavesdropping. 
For qubits, the qubit trine represents the smallest complete set of
non-orthogonal states. 
Earlier trine-based protocols include the schemes by Bechmann-Pasquinucci and
Peres \cite{b.bpp2000} and by Phoenix, Barnett, and Chefles \cite{b.pbc2000}. 

When QKD is performed, it is generally assumed that Alice and Bob share a
common reference frame, the precise nature of which depends on the specific
information carriers involved. 
For instance, correct orientation of Alice and Bob's coordinates is necessary
for proper alignment of the preparation and measurement apparatus. 
Practical implementations of cryptographic protocols require establishing a
rigid shared frame in advance or a frequent automatic realignment.
A lack of a shared reference frame is equivalent to the presence of
decoherence in the quantum channel \cite{b.brs2007}. 

In this contribution, we describe a trine-based cryptographic protocol that uses
reference-frame-free qubits and a novel scheme for key generation.
We report the asymptotic noise threshold below which this QKD procedure 
is secure.
The security analysis involves a very plausible, yet unproven, simplifying
symmetry assumption.

\section{Basics of trine schemes}

\begin{figure}
\centerline{\includegraphics[scale=0.7]{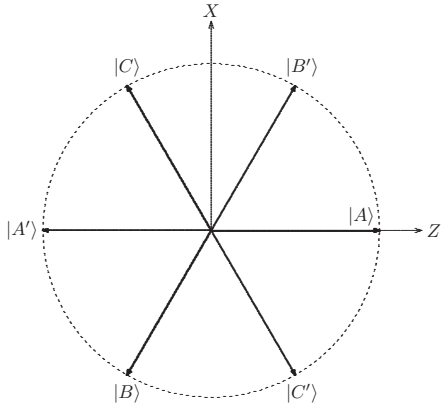}}
\caption{\label{fig.trinebloch}%
Qubit trine $T$ and complementary trine $T'$ in the Bloch representation. 
The projectors on the respective kets are all symbolized by vectors in the
$XZ$-plane.} 
\end{figure}

A qubit trine is represented by a symmetric set of three states lying in the
$XZ$-plane of the Bloch sphere, where adjacent vectors are separated by
$120\degree$; see fig.~\ref{fig.trinebloch}. 
Let us call the qubit trine ${T = \{ \ket{A}, \ket{B}, \ket{C} \}}$. 
Alice prepares her qubits in any one of the trine states with equal
probability, and sends these qubits one at a time to Bob. 
Bob measures the qubits he receives with a probability
operator measurement (POM) whose outcomes are not projectors to the trine $T$,
such as $\proj{A}$ but rather to states orthogonal to $T$, i.e., states
belonging to the set ${T' = \{ \ket{A'}, \ket{B'}, \ket{C'} \}}$, where 
\begin{equation}
\label{eq.trineprop}
\bigl|\bracket{A}{A'}\bigr|^2 = 0\,, 
\quad \bigl|\bracket{A}{B'}\bigr|^2 
    = \bigl|\bracket{A}{C'}\bigr|^2 =\frac{3}{4}\,,
\end{equation}
with analogous relations holding for $\ket{B}$ and $\ket{C}$.
Since Alice works only with the trine $T$ while Bob is concerned only with
states from the complementary trine $T'$, we can simplify matters by treating
corresponding states of $T$ and $T'$ as identical. 
For example, if Alice sends $A$, we say Bob never measures $A$ but has equal
probability of obtaining either $B$ or $C$. 
In this description, the joint probabilities of the quantum communication
channel are given by table~\ref{tab.trinescheme}, for which 
\begin{equation}
\label{eq.trineMI}
I(A:B) = \log_{2}\frac{3}{2} = 0.585
\end{equation}
is the mutual information $I(A:B)$ between Alice and Bob.
With such a noiseless trine channel, then, they can generate
up to $0.585$ secret key bits per qubit sent. 

\begin{table}
\caption{Joint probabilities for the noiseless trine channel.%
\label{tab.trinescheme}}
\centerline{\rule{0pt}{28pt}
\begin{tabular}[t]{r|ccc}
 & A & 
\raisebox{0pt}[0pt]{\begin{tabular}[b]{@{}c@{}}Bob \\ B\end{tabular}}
& C \\ [0.5ex] \hline 
A & 0 & $\displaystyle\frac{1}{6}$ & $\displaystyle\frac{1}{6}$%
\rule{0pt}{18pt}\\
\makebox[0pt][r]{\begin{rotate}{90}\hspace*{-0.8em}Alice\end{rotate}\quad}%
B & $\displaystyle\frac{1}{6}$ & 0 & $\displaystyle\frac{1}{6}$%
\rule{0pt}{18pt}\\ 
C & $\displaystyle\frac{1}{6}$ & $\displaystyle\frac{1}{6}$ & 
0\rule{0pt}{18pt}\\ 
\end{tabular}
}
\end{table}

\section{The double trine scheme}
If Alice and Bob use physical qubits for the logical qubits of the trine, they
must be sure to agree on the coordinates for preparing and measuring
their quantum signals. 
However, they can skip this problem altogether by using reference-frame-free
(RFF) qubits. 
In our scheme a logical qubit is constructed by coupling three physical
qubits. 
For concreteness, we consider the physical qubits to be spin-$\frac{1}{2}$
particles (with $\ket{0}$ for spin-up and $\ket{1}$ for spin-down) 
and combine their angular momenta in the appropriate manner. 

Following the recipe of ref.~\cite{b.ste2008}, we consider two trines.
In the subspace ($j=\frac{1}{2}$, $m=\frac{1}{2}$) we have the trine
 \begin{eqnarray}
\label{eq.pstates}
 \ket{p_1} &=& \bigl(\ket{001} - \ket{010}\bigr)/\sqrt{2}\,, \nonumber\\ 
 \ket{p_2} &=& \bigl(\ket{100} - \ket{001}\bigr)/\sqrt{2}\,, \nonumber\\  
 \ket{p_3} &=& \bigl(\ket{010} - \ket{100}\bigr) /\sqrt{2}\,,
\end{eqnarray}  
and the states
\begin{eqnarray}
\label{eq.qstates}
 \ket{q_1} &=& \bigl(\ket{101} - \ket{110}\bigr)/\sqrt{2}\,,  \nonumber\\ 
 \ket{q_2} &=& \bigl(\ket{110} - \ket{011}\bigr)/\sqrt{2}\,,  \nonumber\\ 
 \ket{q_3} &=& \bigl(\ket{011} - \ket{101}\bigr) /\sqrt{2}\,, 
\end{eqnarray}  
constitute the trine in the subspace ($j=\frac{1}{2}$, $m=-\frac{1}{2}$).
All relevant states are in the ${j=\frac{1}{2}}$ sector of the three
spin-$\frac{1}{2}$ atoms and, for the sake of simplifying the notation, we shall
consistently ignore the empty ${j=\frac{3}{2}}$ sector. 

The sums of the projectors to corresponding $p$- and $q$-states are
\begin{equation}
\label{dts}
W_i=\proj{p_{i}} + \proj{q_{i}}=S_{jk} \,,
\end{equation}
where $S_{jk}$ projects on the singlet sector for atoms $j$ and $k$,
and the indices $ijk$ pertain to all cyclic permutations of $123$.  
By construction, the $W_{i}$s are rotationally invariant and hence have
the same properties for all reference frames---they are RFF operators.
We note that the  
$W_i$s have two eigenvalues $0$ and two eigenvalues $1$, so
that $W_i$ and ${1-W_i}$ project on orthogonal two-dimensional subspaces, and
\begin{equation}
  \label{W-traces}
  \sum_{i=1}^3W_i=\frac{3}{2}\,,\quad
  \tr{W_i}=2\,,\quad\tr{W_iW_j}=\frac{3\delta_{ij}+1}{2}
\end{equation}
are identities that will be relevant in what follows.

Because our scheme has two independent sets of trines, we call it the
\emph{double trine scheme}. 
It works as follows. 
Alice sends a random sequence of the states $\rho_i=\frac{1}{2}W_i$ to Bob,
with the three states occurring with equal frequency, and Bob measures them
with a POM whose outcomes are $\Pi_j=\frac{2}{3}(1-W_j)$.
The resulting joint probabilities,
\begin{equation}
  \label{jointprobs}
  p_{ij}=\frac{1}{3}\tr{\rho_i\Pi_j}=\frac{1-\delta_{ij}}{6}
\end{equation}
are those of table~\ref{tab.trinescheme}.

\section{Signal and idler qubit}
The sum of the three $p$-kets of (\ref{eq.pstates}) vanishes---they are
linearly dependent because the ${j=m=\frac{1}{2}}$ sector is two-dimensional. 
A pair ${\ket{++},\ket{-+}}$ of orthogonal kets is identified by
\begin{equation}
  \label{eq.+kets}
  \bigl(\ket{p_1},\ket{p_2},\ket{p_3}\bigr)
=\bigl(\ket{++},\ket{-+}\bigr)
\frac{1}{\sqrt{2}}\left(
  \begin{array}{rrr}
    1 & \omega^{\phantom{2}} & \omega^2 \\ 1 & \omega^2 & \omega^{\phantom{2}}
  \end{array}\right),
\end{equation}
where $\omega=\exp(\mathrm{i}2\pi/3)$, and likewise we have
\begin{equation}
  \label{eq.-kets}
  \bigl(\ket{q_1},\ket{q_2},\ket{q_3}\bigr)
=\bigl(\ket{+-},\ket{--}\bigr)
\frac{1}{\sqrt{2}}\left(
  \begin{array}{rrr}
    1 & \omega^{\phantom{2}} & \omega^2 \\ 1 & \omega^2 & \omega^{\phantom{2}}
  \end{array}\right)
\end{equation}
for the $q$-states.
We regard the four orthogonal states $\ket{\pm\pm}$ that span the
${j=\frac{1}{2}}$ sectors of the three spin-$\frac{1}{2}$ atoms as two-qubits
states~\cite{b.ste2008} whereby, for example, ket $\ket{+-}$ has the
\emph{signal qubit} in the `$+$' state and the \emph{idler qubit} in the `$-$'
state. 

The signal states ${\ket{A},\ket{B},\ket{C}}$ that we identify as
\begin{equation}
  \label{eq.sigABC}
  \bigl(\ket{A},\ket{B},\ket{C}\bigr)
=\bigl(\ket{+},\ket{-}\bigr)
\frac{1}{\sqrt{2}}\left(
  \begin{array}{rrr}
    1 & \omega^{\phantom{2}} & \omega^2 \\ 1 & \omega^2 & \omega^{\phantom{2}}
  \end{array}\right)  
\end{equation}
form the single-qubit trine that matters.
Upon denoting the Pauli operators of the signal qubit by $X$, $Y$, and $Z$ and
identifying $\ket{\pm}$ with the eigenkets of $Y=\mathrm{i}XZ$, the
signal-qubit trine is in the $XZ$ plane as depicted in
fig.~\ref{fig.trinebloch}. 
 
In view of
\begin{equation}
  \label{eq.sigW}
  W_1=\proj{A}\otimes 1\,,\quad
  W_2=\proj{B}\otimes 1\,,\quad
  W_3=\proj{C}\otimes 1\,,
\end{equation}
the idler sector is completely irrelevant: Alice encodes the information in
the signal qubit only, and Bob's POM does not probe the idler qubit at all.
The sole purpose of the idler qubit is to render possible the
construction of the rotationally invariant signal qubit.
We can, therefore, think of the double trine scheme as a generic scheme of the
kind described in the context of fig.~\ref{fig.trinebloch} with the signal
qubit carrying the quantum state from Alice to Bob.

\section{Common source scenario}
Rather than having Alice prepare qubits in the trine states and send
them to Bob, who then analyzes them with the trine POM, we can generate the
joint probabilities of table~\ref{tab.trinescheme} in a more symmetric and
largely equivalent way.
In this alternative scenario, a source distributes entangled two-qubit states 
to Alice and Bob. 

Ideally, the two signal qubits are in their singlet state that is described
by the statistical operator 
\begin{equation}
\label{eq.sourcestate}
\rho_0=\proj{s}\quad\mbox{with}\enskip
\ket{s} = \frac{\ket{+-}-\ket{-+}}{\sqrt{2}}\,.
\end{equation}
On their respective qubits, Alice and Bob then both measure the same trine POM
with the outcomes 
\begin{equation}
\label{eq.trinepovm}
\Pi_{i} = \proj[\frac{2}{3}]{i}\quad\mbox{for}\enskip i=A,B,C\,.
\end{equation}
Indeed, the resulting joint probabilities,
\begin{equation}
\label{eq.jointprob}
p_{jk}= \tr{\Pi_{j}\otimes \Pi_{k}\,\rho_0} \quad\mbox{for}\enskip
j,k = A,B,C\,,
\end{equation}
are those of table~\ref{tab.trinescheme}.

In the security analysis below, we shall assume that the source is controlled
by eavesdropper Eve.
Her activities will introduce noise into the quantum channel between Alice and
Bob, but they are only accepting qubits from a source that \emph{looks like}
the singlet of (\ref{eq.sourcestate}) with an admixture of unbiased noise,
\begin{equation}
\label{eq.unbiasednoise}
\rho_\epsilon = \proj[(1-\epsilon)]{s} + \frac{\epsilon}{4}
\end{equation}  
with $0 \leq \epsilon \leq 1$.

\begin{table}
\caption{Joint probabilities for the noisy trine channel.%
\label{tab.noisytrine}}
\centerline{\rule{0pt}{28pt}
\begin{tabular}[t]{r|ccc}
 & A & 
\raisebox{0pt}[0pt]{\begin{tabular}[b]{@{}c@{}}Bob \\ B\end{tabular}}
& C \\ [0.5ex] \hline 
A & $\displaystyle\frac{\epsilon}{9}$ 
  & $\displaystyle\frac{3-\epsilon}{18}$ 
  & $\displaystyle\frac{3-\epsilon}{18}$%
\rule{0pt}{18pt}\\
\makebox[0pt][r]{\begin{rotate}{90}\hspace*{-0.8em}Alice\end{rotate}\quad}%
B & $\displaystyle\frac{3-\epsilon}{18}$ 
  & $\displaystyle\frac{\epsilon}{9}$ 
  & $\displaystyle\frac{3-\epsilon}{18}$%
\rule{0pt}{18pt}\\ 
C & $\displaystyle\frac{3-\epsilon}{18}$
  & $\displaystyle\frac{3-\epsilon}{18}$
  & $\displaystyle\frac{\epsilon}{9}$\rule{0pt}{18pt}\\ 
\end{tabular}
}
\end{table}

As far as Alice and Bob are concerned, the noise parameter $\epsilon$
characterizes the channel.  
In the presence of noise, they observe errors in the trine channel:
sometimes they get the same measurement outcome for a particular
qubit pair, which does not happen in the noise-free case.
Rather than the noise-free joint probabilities of table~\ref{tab.trinescheme},
they now have the probabilities of table~\ref{tab.noisytrine}.

But since their measurements yield only these nine joint probabilities, 
Alice and Bob cannot determine all fifteen parameters that specify the
two-qubit state distributed by the source.
In this respect, the trine schemes are markedly different from tomographic
protocols~\cite{b.tomocrypt}, such as the six-state
protocol~\cite{b.sixstates} or the Singapore protocol~\cite{b.SingProt}, in
which full tomography of the source state is central.

If Alice and Bob do not see the symmetric probability
table~\ref{tab.noisytrine}, they enforce the symmetry by twirling.
For this purpose, they carry out random bilateral rotations on the qubits that
leave the singlet component intact while removing any bias from the noise.

\section{Efficient generation of the raw dual key}
Once Alice and Bob finish collecting and measuring their qubits, they get a
paired record of measurement results. 
Next, they communicate over an authenticated public channel to discuss the raw
data and distill a cryptographic key. 
Here we describe a new key generation method that yields mutual information
between Alice and Bob closer to the Shannon limit for a trine-based channel
\cite{b.chua2006}. 
We illustrate the procedure with the sample results shown in
table~\ref{tab.sample}. 

\begin{table}
\caption{Example of measurement records for Alice and Bob.%
\label{tab.sample}}
\centerline{\rule{0pt}{28pt}
\begin{tabular}{lccccccr}
      & 1 & 2 & 3 & 4 & 5 & 6 & 7 \\
Alice & A & C & C & B & B & A & C \\
Bob   & B & A & A & C & A & C & B	
\end{tabular}
}
\end{table}

To begin, Alice chooses two time slots in a specified order where her outcomes
are different. 
Suppose she selects columns 2 and 5 in table~\ref{tab.sample}.
Alice's pair of letters in these positions is CB.
She tells Bob to look at his record at those two particular time slots and he
finds he has A in both. 
He declares he has the same letter in both positions. 
Alice quickly determines this letter to be A since it is the only result
consistent with the expected outcomes for a trine protocol. 
Both record A for the key. 
Because there are three possibilities in this scenario, we call this the trit
key. 
Alice and Bob discard the used time slots.

There is another situation to consider. 
Say for the next round, Alice chooses columns 1 and 4. 
Bob finds BC for these time slots and
announces the following: Record $0$ for BC and $1$ for CB. 
Since Alice has AB, she infers that Bob must have BC, and 
both of them record $0$ for the key. 
In this case, there are two possibilities, so we call it the bit key. 
Note that the order of the time slots selected matters: they would both record
$1$ if Alice reversed the order. 

The two situations---the \emph{trit case} and the \emph{bit case}---are 
mutually exclusive events so
the bit and trit keys are independently built up from the raw data.
In the noiseless case, the trit case happens $\frac{1}{4}$ of the time
while the bit case happens in the remaining $\frac{3}{4}$. 
It follows that the number of key bits, per qubit exchanged, 
that Alice and Bob share in the key sequences thus generated is given by
\begin{equation}
\label{eq.mutinfotrine}
I(A:B) = \frac{1}{2} \left( \frac{1}{4} \log_{2} 3 
+ \frac{3}{4} \log_{2} 2\right)= 0.573\,, 
\end{equation}  
which is 98\% of the Shannon limit in (\ref{eq.trineMI}).
 
Noise in the channel leads to errors in the shared keys, since the unexpected
result of getting the same letter during transmission will sometimes occur.
The probability that the next letter pair contributes an entry to the trit key
is now 
${p_\mathrm{trit}=\frac{1}{12}(3-\epsilon)(1+\epsilon)}$, and the probability of
contributing to the bit key is 
${p_\mathrm{bit}=\frac{1}{12}[(3-\epsilon)^2+4\epsilon]}$.
 
In the trit case, the correctly matched pairs in the key (that is, both Alice
and Bob write down the same letter, whether A, B, or C) each have probability
$(3-\epsilon)/(9+9\epsilon)$;
the other six outcomes where they disagree have probability
$2\epsilon/(9+9\epsilon)$ each. 
Likewise in the bit case, the two instances when Alice and Bob agree both have
probability $\frac{1}{2}(3-\epsilon)^{2}/[(3-\epsilon)^2+4\epsilon]$,
while for the other two where they disagree the probability is
$2\epsilon/[(3-\epsilon)^2+4\epsilon]$ each. 

These probabilities yield
\begin{eqnarray}
\label{eq.noisymutinfotrine}
\nonumber I_{\mathrm{trit}}(A:B) 
&=&  
\frac{3-\epsilon}{3+3\epsilon} \log_{2} \frac{3-\epsilon}{1+\epsilon} 
\\ && + 
\frac{4\epsilon}{3+3\epsilon} \log_{2} \frac{2\epsilon}{1+\epsilon}\,, 
\nonumber\\ 
\nonumber I_{\mathrm{bit}}(A:B) &=&  
\frac{(3-\epsilon)^{2}}{(3-\epsilon)^{2}+4\epsilon} 
\log_{2} \frac{2(3-\epsilon)^{2}}{(3-\epsilon)^{2}+4\epsilon}  
\\  & & + 
\frac{4\epsilon}{(3-\epsilon)^{2}+4\epsilon}  
\log_{2} \frac{8\epsilon}{(3-\epsilon)^{2}+4\epsilon}\qquad  
\end{eqnarray} 
for the resulting mutual information between Alice and Bob for the two key
sequences. 
For $\epsilon=0.1$, their weighted sum 
${\frac{1}{2}(p_\mathrm{trit}I_{\mathrm{trit}}%
+p_\mathrm{bit}I_{\mathrm{bit}})}$
equals 96.4\% of the Shannon limit, the
mutual information of the joint probabilities in table~\ref{tab.noisytrine}.
As functions of $\epsilon$, $I_{\mathrm{bit}}(A:B)$ and
$I_{\mathrm{trit}}(A:B)$ are the monotonously decreasing curves in
figs.~\ref{fig.twobit} and \ref{fig.twotrit} below, respectively.

\section{Security analysis}
Eve is given full control of the source and is allowed to keep a quantum
record, encoded in ancilla states, of what is sent.
We write the source state in the form
\begin{equation}
\label{eq.sourceanc}
\ket{S}=\ket{++\;E_1}+\ket{+-\;E_2}+\ket{-+\;E_3}+\ket{--\;E_4}\,,
\end{equation}
where, for example, $\ket{+-\;E_2}$ is the `$+$' state of (\ref{eq.sigABC})
for Alice's signal qubit, the `$-$' state for Bob's, and Eve's ancilla in
state $\ket{E_2}$. 
When Alice's POM gives the $j$th outcome, and Bob's the $k$th, the reduced
ancilla state is described by $\ket{E_{jk}}$ where $j$ and $k$ independently
take on values of $A$, $B$, or $C$.
After accounting for the coefficients in (\ref{eq.sigABC}) and
(\ref{eq.trinepovm}), we have
\begin{equation}
  \label{eq.anc-jk}
  \ket{E_{jk}}=\bigl(\ket{E_1},\ket{E_2},\ket{E_3},\ket{E_4}\bigl)
  \,\frac{1}{3}{\left(\begin{array}{c}
   \omega^{-j-k}\\ \omega^{-j+k} \\ \omega^{j-k} \\ \omega^{j+k}
  \end{array}\right)}
\end{equation}
with $ABC\widehat{=}012$ for the $j$ and $k$ values in the exponents. 

The joint probabilities of table~\ref{tab.noisytrine} impose the constraints
\begin{equation}
  \label{eq.Eab-constr}
  p_{jk}=\bracket{E_{jk}}{E_{jk}}=\frac{\epsilon}{9}\delta_{jk}
                               +\frac{3-\epsilon}{18}(1-\delta_{jk})\,,
\end{equation}
which in turn imply 
\begin{eqnarray}
\label{eq.Eiproplist}
&&
\bracket{E_1}{E_1}+\bracket{E_2}{E_2}+\bracket{E_3}{E_3}+\bracket{E_4}{E_4}=1\,,
\nonumber\\
&&\bracket{E_{1}}{E_{2}} + \bracket{E_{3}}{E_{4}} = 0
=\bracket{E_{1}}{E_{3}} + \bracket{E_{2}}{E_{4}}\,, \nonumber\\
&&\bracket{E_{1}}{E_{4}} = 0\,,\qquad
\bracket{E_{2}}{E_{3}} =-(1-\epsilon)/2\,.
\end{eqnarray}
These determine nine of the 16 real parameters that specify the positive 
${4\times4}$ matrix of the $\bracket{E_j}{E_k}$ amplitudes. 

A convenient choice of the remaining seven real parameters is given by
representing the kets $\ket{E_1}$, \dots, $\ket{E_4}$ by the columns of a
matrix of the form~\cite{b.tabia2009} 
\begin{equation}
\label{eq.4x4-V}
V = \left(
\begin{array}{c@{\quad}c@{\quad}c@{\quad}c}
a^{\ }_1 & \lambda a^{\ }_2 & -\mu a^{\ }_2 & 0 \\
0 & r^{\ }_1\cos\theta & -r^{\ }_2\mathrm{e}^{\mathrm{i}\phi}\sin\theta & 0 \\
0 & r^{\ }_1\mathrm{e}^{-\mathrm{i}\phi}\sin\theta & -r^{\ }_2\cos\theta & 0 \\
0 & \mu^{*} a^{\ }_1 & -\lambda^{*} a^{\ }_1 & a^{\ }_2
\end{array}
\right),
\end{equation}
where $a^{\ }_1,a^{\ }_2,r^{\ }_1,r^{\ }_2,\phi,\theta$ are real and
$\lambda,\mu$ are complex, and their values are subject to 
\begin{eqnarray}
  \label{eq.v-constr}
  \bigl(1+|\lambda|^2+|\mu|^2\bigr)\bigl(a_1^2+a_2^2\bigr)
     +r_1^2+r_2^2&=&1\,,\nonumber\\
  \lambda^*\mu\bigl(a_1^2+a_2^2\bigr)
  +r^{\ }_1r^{\ }_2\mathrm{e}^{\mathrm{i}\phi}\sin(2\theta)
  &=&\frac{1-\epsilon}{2}\,.
\end{eqnarray}
As demonstrated by 
\begin{eqnarray}
  \label{eq.rho-eps}
  &&a^{\ }_1=a^{\ }_2=\frac{1}{2}\sqrt{\epsilon}\,,\quad
    r^{\ }_1=r^{\ }_2=\frac{1}{2}\sqrt{2-\epsilon}\,,\quad
    \lambda=\mu=0\,,\nonumber\\
  &&\phi=0\,,\quad\sin(2\theta)=\frac{2-2\epsilon}{2-\epsilon}\,,
\end{eqnarray}
for which Alice and Bob's reduced two-qubit state is $\rho_{\epsilon}$ of
(\ref{eq.unbiasednoise}), there surely are permissible values, but it is 
not obvious which set of parameters is optimal for Eve.
With the aid of (\ref{eq.anc-jk}), each permissible $V$ matrix gives us valid
column representations for the $\ket{E_{jk}}$s.

In fact, Eve is not interested in distinguishing the $\ket{E_{jk}}$
states themselves but rather the two-ancilla states that are associated with 
symbols in the key sequences, whereby the bit and trit cases need to be
considered separately. 

In the \emph{bit case}, the two-ancilla state conditioned on Alice concluding
that Bob has the letter sequence `$jk$' is given by
\begin{eqnarray}
 \rho _{jk}^{(A)} &\propto& \proj{E_{kj}E_{lk}} + \proj{E_{lj}E_{jk}} 
\nonumber\\&& + \proj{E_{kj}E_{jk}} + \proj{E_{kk}E_{lj}}
\nonumber\\ && + \proj{E_{lk}E_{jj}} + \proj{E_{kk}E_{jj}}\,,
\end{eqnarray}
where $jkl$ can be any permutation of $ABC$.  
The first three terms account for the cases in which Alice and Bob record the
same bit value, and the bit errors are covered by the last three terms. 
For example, the first term is for the situation when Alice has `$kl$' and Bob
has `$jk$' while both have `$kj$' for the last term.
Eve has to tell $\rho _{jk}^{(A)}$ and $\rho _{kj}^{(A)}$ apart when Bob
announces that his letters are `$j$' and `$k$'.

Analogously, in the \emph{trit case}, we have the conditioned two-ancilla state
\begin{eqnarray}
\rho_{j}^{(A)} &\propto&\proj{E_{kj}E_{lj}} + \proj{E_{lj}E_{kj}} 
\nonumber\\ &&+ \proj{E_{kk}E_{lk}} + \proj{E_{lk}E_{kk}} 
\nonumber\\ &&+ \proj{E_{kl}E_{ll}} + \proj{E_{ll}E_{kl}}
\end{eqnarray}
when Alice concludes that Bob has letter `$j$' twice,
with the first two terms accounting for a correct assignment and the remaining
four terms for errors.
Here, too,  $jkl$ is a permutation of $ABC$, and Eve need to distinguish the
three states $\rho_A^{(A)}$,  $\rho_B^{(A)}$, and $\rho_C^{(A)}$.

The six $\rho_{jk}^{(A)}$s and three $\rho_{j}^{(A)}$s account for 54 of the
81 two-ancilla kets $\ket{E_{jk}E_{j'k'}}$. 
This is as it should be because the remaining 27 kets are those for which
Alice has the same letter twice, and this situation does not occur.

If Eve eavesdrops on Bob, the conditioned two-ancilla states are different. 
In the bit case we have  
\begin{eqnarray}
\rho _{jk}^{(B)} &\propto& \proj{E_{kj}E_{lk}} + \proj{E_{lj}E_{jk}} 
\nonumber\\ &&+ \proj{E_{kj}E_{jk}} + \proj{E_{lj}E_{kk}} 
\nonumber\\ &&+ \proj{E_{jj}E_{lk}} + \proj{E_{jj}E_{kk}}\,,  
\end{eqnarray}
and the states 
\begin{eqnarray}
\rho _{j}^{(B)} &\propto& \proj{E_{kj}E_{lj}} + \proj{E_{lj}E_{kj}} 
\nonumber\\ &&+ \proj{E_{jj}E_{lj}} + \proj{E_{lj}E_{jj}} 
\nonumber\\ &&+ \proj{E_{jj}E_{kj}} + \proj{E_{kj}E_{jj}}
\end{eqnarray}
apply in the trit case.
They differ from their respective counterparts by the error terms. 
Therefore, we explore both sets of ancilla states to see whether Eve gains any
advantage by eavesdropping on either Alice or Bob, or if it does not make any
difference to the optimal amount of information she can obtain. 

With the assignment of signal-qubit Pauli operators $X$, $Y$, $Z$ discussed
above in the context of (\ref{eq.sigABC}), the two-qubit state $\rho_{AB}^{\ }$
that the source distributes to Alice and Bob is specified by the eight fixed
expectations values 
\begin{eqnarray}
  \label{eq.ABexpect1}
&&\expect{X_A}=\expect{Z_A}=\expect{X_B}=\expect{Z_B}=0\,,\nonumber\\
&&\expect{X_AX_B}=\expect{Z_AZ_B}=-(1-\epsilon)\,,\nonumber\\
&&\expect{X_AZ_B}=\expect{Z_AX_B}=0
\end{eqnarray}
together with the seven adjustable expectation values
\begin{eqnarray}
  \label{eq.ABexpect2}
&&\frac{1}{2}\bigl(\expect{Y_A}\pm\expect{Y_B}\bigr)=\left\{
  \begin{array}{l}
    a_1^2-a_2^2\,,\\[1ex]
    r_1^2-r_2^2-(|\lambda|^2-|\mu|^2)(a_1^2-a_2^2)\,,
  \end{array}\right.\nonumber\\
&&\expect{Y_AZ_B}+\mathrm{i}\expect{Y_AX_B}=4\lambda a_1a_2\,,\nonumber\\
&&\expect{Z_AY_B}+\mathrm{i}\expect{X_AY_B}=-4\mu a_1a_2\,,\nonumber\\
&&\expect{Y_AY_B}=2(a_1^2+a_2^2)-1\,,
\end{eqnarray}
which reveal the physical significance of the seven free parameters in
(\ref{eq.4x4-V}). 
Alice and Bob cannot distinguish between $\rho_{AB}^{\ }$, 
$X_AX_B\rho_{AB}^{\ }X_AX_B$, $Y_AY_B\rho_{AB}^{\ }Y_AY_B$, and
$Z_AZ_B\rho_{AB}^{\ }Z_AZ_B$, and Eve gets the same amount of information
from the corresponding four sets of conditioned ancilla states.
It follows that Eve can just as well choose the parameters in (\ref{eq.4x4-V})
such that
$\rho_{AB}^{\ }={X_AX_B\rho_{AB}^{\ }X_AX_B}={Z_AZ_B\rho_{AB}^{\ }Z_AZ_B}$.
Then, the six expectation values in (\ref{eq.ABexpect2}) that involve a
single $Y$ vanish, which happens for 
\begin{equation}\label{eq.noYs}
a^{\ }_{1}= a^{\ }_{2}\,, \quad r^{\ }_{1} = r^{\ }_{2}\,,\quad
\lambda =\mu =0\,.
\end{equation}
Indeed, it is plausible, and supported by much numerical evidence, that a
parameter choice that yields such a particularly noisy  $\rho_{AB}^{\ }$ is
advantageous for Eve because then the entanglement between her ancilla and the
qubits for Alice and Bob is particularly strong.

With (\ref{eq.noYs}), matrix $V$ takes on the simple one-parameter form
\begin{equation}
\label{eq.c-V}
V = \frac{1}{2}\left(
\begin{array}{crrc}
\sqrt{c} & 0 & 0 & 0 \\
0 & \phantom{-}x & -y & 0 \\
0 & y & -x & 0 \\
0 & 0 & 0 & \sqrt{c}
\end{array}
\right)
\end{equation}
with ${0\leq c\leq 2\epsilon}$ and ${x\pm y =\sqrt{2-c\pm2(1-\epsilon)}}$.
We return to (\ref{eq.rho-eps}) for ${c=\epsilon}$, while ${c=2\epsilon}$ and
${c=2\epsilon-\epsilon^2}$ give the $\rho_{AB}^{\ }$s with minimal concurrence
and maximal entropy, respectively; the $\rho_{AB}^{\ }$s for
${2\epsilon+c\geq2}$ are separable \cite{b.ase2006}. 

The following observation lends additional support to (\ref{eq.noYs}) and
(\ref{eq.c-V}):
The resulting conditioned ancilla states are such that it does not matter
which letter pairs `$jk$' and `$kj$' are to be distinguished in the bit case,
or which letter `$j$' is the actual one in the trit case. 
Eve does not acquire better knowledge about a subset of key entries at the
price of knowing less about other subsets. 
By contrast, such an asymmetry in her knowledge is typically the case if some
of the single-$Y$ expectation values in (\ref{eq.ABexpect2}) are nonzero.

\begin{figure}
\centerline{\includegraphics{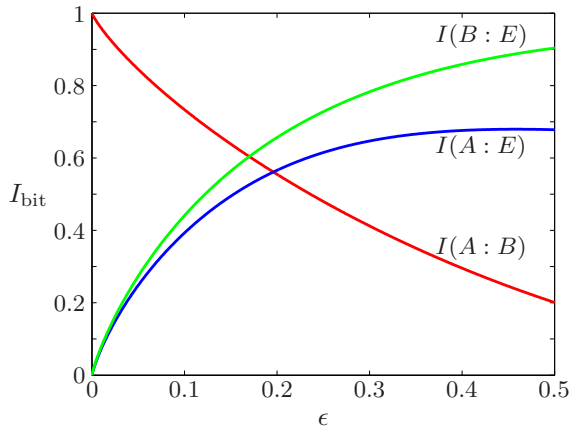}}
\caption{Optimizing the one-parameter source state: 
Wiretapper bound for Eve eavesdropping on the bit key. 
For Alice, the noise threshold is $\epsilon = 0.197$, where 
$I_\mathrm{bit} = 0.560$. 
The corresponding numbers for Bob are $\epsilon = 0.170$, 
$I_\mathrm{bit} = 0.603$.}
\label{fig.twobit}
\end{figure}

\begin{figure}
\centerline{\includegraphics{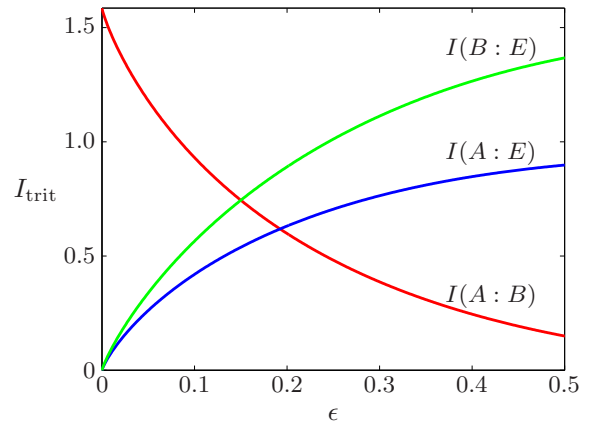}}
\caption{Optimizing the one-parameter source state: 
Wiretapper bound for Eve eavesdropping on the trit key. 
For Alice, the noise threshold is $\epsilon = 0.193$, where
$I_\mathrm{trit} = 0.618$. 
The corresponding numbers for Bob are $\epsilon =0.150$, 
$I_\mathrm{trit} = 0.744$.} 
\label{fig.twotrit}
\end{figure}

Accepting thus the hypothesis that it suffices to consider
matrices $V$ of the single-parameter form (\ref{eq.c-V}), we take the
resulting two-ancilla kets $\ket{E_{jk}}$ and calculate the 
Holevo-Schumacher-Westmoreland (HSW) bounds \cite{b.hol1973,b.schumwest1997}
on ${I(A:E)}$ and ${I(B:E)}$ as a function of $c$. 
After optimizing the value of $c$ for the given value of the noise parameter
$\epsilon$, we obtain the monotonically increasing
curves in figs.~\ref{fig.twobit} and
\ref{fig.twotrit} for the bit key and the trit key, respectively.

The $\epsilon$ values for which these curves intersect the curves representing
the corresponding $I(A:B)$ of (\ref{eq.noisymutinfotrine}) determine the noise
thresholds below which Alice and Bob can generate a secret key from the raw
key by the usual procedures of error correction and privacy amplification
\cite{b.renner2008}.  
Both in the bit case and in the trit case, the thresholds are higher when Eve
is eavesdropping on Alice than on Bob.
We could not find lower thresholds with any parameter values not restricted by
the symmetry requirements (\ref{eq.noYs}).

\section{Summary and discussion}
We described a basis-independent trine protocol for QKD that uses RFF signal
qubits encoded in mixed states of three physical qubits.
The protocol exploits a novel efficient key generation scheme that yields a
dual alphabet key. 
We analyzed the security with a plausible symmetry assumption that simplifies
the task to the optimization of a single parameter.
As a consequence of the asymmetric roles played by them during the
key generation, there are different noise thresholds for
eavesdropping on Alice and Bob.

The raw keys need to be processed before Alice and Bob share a secret key.
For the error correction and the privacy amplification one of the raw keys
serves as the error-free reference, and we choose Alice's key for this purpose 
because then the higher thresholds apply.
We conclude that a secret key can be generated for ${\epsilon<0.197}$, and one
should stay well below this threshold to have a good key bit rate.

Regarding practical implementations of the scheme, we note that the production 
of entangled states is no routine matter, with the difficulty
increasing rapidly with size. 
A practical system with a common source for Alice and Bob requires six
entangled physical qubits for each transmission.
It is easier to use the variant where Alice prepares the states and
sends them to Bob as this requires only three qubits.
As with other QKD protocols, photon polarization is the most likely
candidate for the physical qubits. 
Alice could prepare three-photon trine states by first preparing two of the
three photons in a Bell state and the third photon with random
polarization; it is possible to achieve this by beginning with an entangled
four-photon state and measuring the polarization of the fourth photon with a
suitable POM.  
Bob's POM would then test if one of the three orthogonal Bell states is
present for every trio of photons received from Alice.
Given the limited efficiency of typical photodetectors, efficient detection of
all three photons is a challenge though. 

\acknowledgments
We are grateful for useful discussions with Jun Suzuki, Syed~M. Assad,
and Valerio Scarani.
BGE thanks Hans Briegel for the kind hospitality in Innsbruck where part of
this work was done. 
Centre for Quantum Technologies is a Research Centre for Excellence funded by
the Ministry of Education and National Research Foundation of Singapore.

\end{document}